\documentclass[reprint,aps,prl,superscriptaddress,amsmath,amssymb,floatfix,a4paper]{revtex4-1}
\usepackage{graphicx}
\usepackage{layout}
\usepackage{dcolumn}
\usepackage{braket}
\usepackage{bm}
\usepackage{color}
\usepackage{verbatim}
\usepackage{txfonts}

\begin{document}

\title{Antilocalization of Coulomb Blockade in a Ge-Si Nanowire}

\author{A. P. Higginbotham}
\affiliation{Center for Quantum Devices, Niels Bohr Institute, University of Copenhagen, 2100 Copenhagen, Denmark}
\affiliation{Department of Physics, Harvard University, Cambridge, Massachusetts, 02138, USA}

\author{F. Kuemmeth}
\affiliation{Center for Quantum Devices, Niels Bohr Institute, University of Copenhagen, 2100 Copenhagen, Denmark}

\author{T. W. Larsen}
\affiliation{Center for Quantum Devices, Niels Bohr Institute, University of Copenhagen, 2100 Copenhagen, Denmark}

\author{M. Fitzpatrick}
\affiliation{Center for Quantum Devices, Niels Bohr Institute, University of Copenhagen, 2100 Copenhagen, Denmark}
\affiliation{Department of Physics, Middlebury College, Middlebury, Vermont 05753, USA}

\author{J. Yao}
\affiliation{Department of Chemistry and Chemical Biology, Harvard University, Cambridge, Massachusetts 02138, USA}

\author{H. Yan}
\affiliation{Department of Chemistry and Chemical Biology, Harvard University, Cambridge, Massachusetts 02138, USA}

\author{C. M. Lieber}
\affiliation{Department of Chemistry and Chemical Biology, Harvard University, Cambridge, Massachusetts 02138, USA}
\affiliation{School of Engineering and Applied Sciences, Harvard University, Cambridge, Massachusetts 02138, USA}

\author{C. M. Marcus}
\affiliation{Center for Quantum Devices, Niels Bohr Institute, University of Copenhagen, 2100 Copenhagen, Denmark}


\begin{abstract}
The distribution of Coulomb blockade peak heights as a function of magnetic field is investigated experimentally in a Ge-Si nanowire quantum dot. Strong spin-orbit coupling in this hole-gas system leads to antilocalization of Coulomb blockade peaks, consistent with theory. In particular, the peak height distribution has its maximum away from zero at zero magnetic field, with an average that decreases with increasing field. Magnetoconductance in the open-wire regime places a bound on the spin-orbit length ($l_{\mathrm{so}} < 20~\mathrm{nm}$), consistent with values extracted in the Coulomb blockade regime ($l_{\mathrm{so}} < 25~\mathrm{nm}$).
\end{abstract}


\maketitle

Antilocalization, a positive correction to classical conductivity, is commonly observed in mesoscopic conductors with strong spin-orbit coupling \cite{Hikami:1980wh,Bergmann:1982vo}, and has been well studied in low-dimensional systems over the past two decades \cite{Bergmann:1984ty,Moyle:1987iz,Dresselhaus:1992eo,Knap:1996en,Kallaher:2010hu,He:2011ix}. In quantum wires (1D) and dots (0D), the combination of coherence and spin-orbit coupling is a topic of renewed interest in part due to numerous quantum information processing proposals---from spin qubits to Majorana modes---where these ingredients play a fundamental role \cite{DanielLoss:1998ia,Zutic:2004fo,Lutchyn:2010hp,Oreg:2010gk}. Antilocalization in 1D systems has been  investigated in detail both theoretically \cite{Beenakker:1988dv,Kurdak:1992cp,Kettemann:2007eg} and experimentally \cite{Hansen:2005ek,Lehnen:2007ey,EstevezHernandez:2010du,Kallaher:2010iq}.
In 0D systems, antilocalization in both the opened and nearly-isolated Coulomb blockade regime has been studied theoretically \cite{Aleiner:2001fn,Ahmadian:2006jx}, but to date experiments have only addressed the open-transport regime, where Coulomb effects play a minor role \cite{Zumbuhl:2002ez,Hackens:2002ei}.

The hole gas formed in the Ge core of a Ge-Si core-shell nanowire \cite{Lu:2005jx} is an attractive system for exploring the coexisting effects of coherence, confinement, and spin-orbit coupling. 
Tunable quantum dots have been demonstrated in this system \cite{Hu:2007hu,Roddaro:2008jc}, band structure calculations indicate strong spin-orbit coupling \cite{Kloeffel:2011bg}, and antilocalization has been demonstrated in the open-transport regime \cite{Hao:2010di}.

In this Letter, we investigate full distributions of Coulomb blockade peak height as a function of magnetic field in a gated Ge-Si core-shell nanowire.  The low-field distributions are consistent with random matrix theory (RMT) \cite{Ahmadian:2006jx} of Coulomb blockade transport through a 0D system with symplectic symmetry (valid for strong spin-orbit coupling), and inconsistent with predictions for orthogonal symmetry (low spin-orbit coupling).  The high-field peak height distribution is found to be a scaled version of the low-field distribution, as expected from theory. However, the observed scale factor, $\sim2.3$, is significantly larger than the theoretical factor of 1.4 \cite{Ahmadian:2006jx}.
Temperature dependence of the peak-height variance is consistent with theory using a value for orbital level spacing measured independently via Coulomb blockade spectroscopy. Consistent bounds on the spin orbit length, $l_{\mathrm{so}} \lesssim 20-25~\mathrm{nm}$, are found in the Coulomb blockade and open transport regimes.

\begin{figure}[b]
\includegraphics{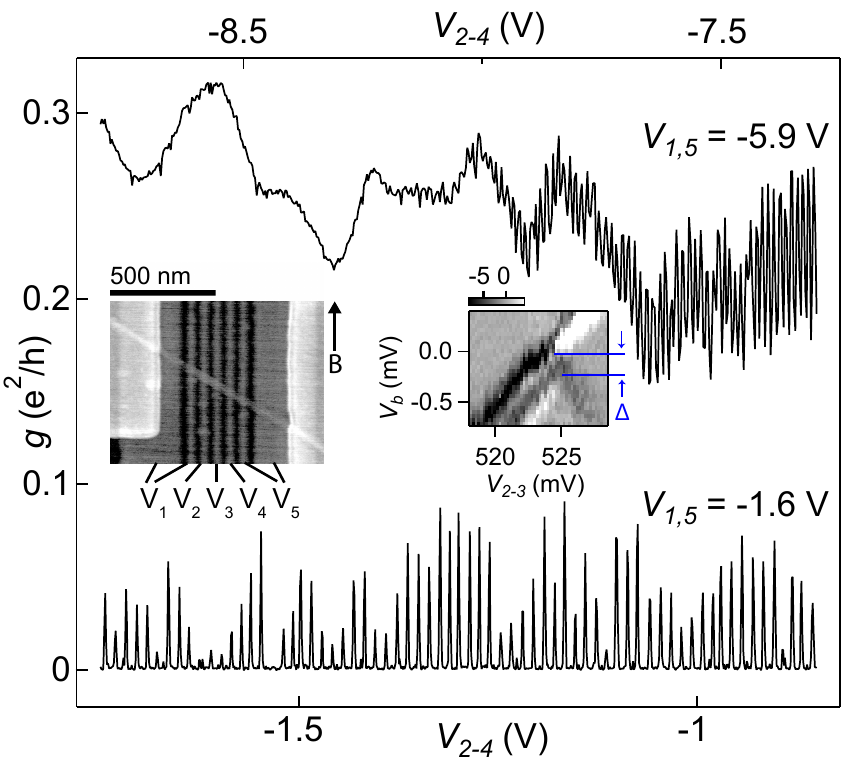}
\caption{\label{fig:1}
Device conductance, $g$, as a function of gate voltages, $V_{2-4}$ [notation indicates $V_{2}=V_{3}=V_{4}$], with $V_{ac} = 50~\mathrm{\mu V}$.
The device can be configured as an open wire (top trace), or an isolated quantum dot (bottom trace).
\textit{Left inset:} SEM micrograph of lithographically identical device. Direction of magnetic field $B$ indicated by vertical arrow.
\textit{Right inset:} $dg/dV_{B}$ in $(\mathrm{e}^2/h)/\mathrm{mV}$ as a function of dc bias, $V_{b}$, and gate voltages, $V_{2-3}$, yield orbital energy spacing $\Delta\sim0.2~\mathrm{meV}$.
}
\end{figure}

Coulomb blockade of transport through a 0D system occurs when temperature, voltage bias, and lifetime broadening are small compared to the charging energy,  $k T, V, \, \Gamma \ll e^2/C$, where $V = V_{b}+V_{ac}$ is the (dc + ac) bias across the device, $C$ is the dot capacitance, $e$ is the electron charge, and  $\Gamma / \hbar = \tau_{escape}^{-1}$ is the tunnel rate out of the system.
When, in addition, $kT$, $V$, and $\Gamma$ are less than the orbital level spacing, $\Delta$, tunneling occurs through a single (ground-state) wave function. In this latter case, Coulomb blockade conductance peaks fluctuate in height from peak to peak, depending on the coupling of the ground-state wave function to modes in the leads.

\begin{figure}[t]
\includegraphics{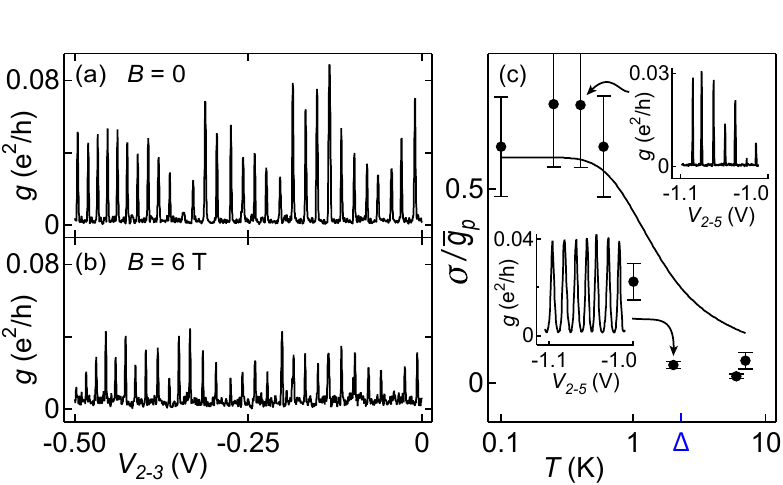}
\caption{\label{fig:2} 
Conductance, $g$, as a function of gate voltages $V_{2-3}$ with the device configured as a quantum dot for (a) $B=0$ and (b) $B=6 \mathrm{T}$.
The application of a magnetic field reduces the average peak height.
(c) Peak height standard deviation normalized by the ensemble-averaged peak height, $\sigma/\overline{g}_p$, versus temperature $T$ at $B=0$ ($V_{ac}=10~\mathrm{\mu V}$), based on $\sim$ 50 peaks per point.
Fluctuations of peak heights decrease for $kT \sim \Delta$. Theory curve has no free parameters (see text). Insets: sample of peaks showing diminished fluctuations at higher temperature.}
\end{figure}

For disordered or chaotic dots, the statistics of these couplings can be calculated from random matrix ensembles for the dot Hamiltonian: orthogonal ($\beta=1$) for time-reversal symmetric systems, unitary ($\beta=2$) for systems with broken time-reversal symmetry, and symplectic ($\beta=4$) for time-reversal symmetric systems with broken spin rotation symmetry. Including spin-orbit and Zeeman coupling yields an extended random matrix theory with two more parameters, $s$ and $\Sigma$, in addition to the usual Dyson parameter,  $\beta$ \cite{Aleiner:2001fn}. The parameter $s$, reflecting Kramers degeneracy, decreases from 2 to 1 with sufficient applied magnetic field; the parameter $\Sigma$, reflecting the mixing of Kramers-split levels, increases from 1 (unmixed) to 2 (mixed) with a sufficient combination of spin-orbit coupling and magnetic field. These ensembles have been investigated experimentally in the open-transport regime \cite{Zumbuhl:2005ix}. 

Writing $\Gamma = \Gamma_{l} + \Gamma_{r}$ for left and right leads, the Coulomb peak height for $\Gamma \ll kT$ is given by 
\begin{equation}
\label{eq:gmax}
g_{p} = \frac{2 e^2}{ \hbar } \frac{ \chi_s }{ k T } \frac{ \Gamma_{l} \Gamma_{r} }{ \Gamma_{l} + \Gamma_{r} }
= \frac{e^2}{\hbar} \frac{ \overline{ \Gamma} }{ 2 k T } \chi_s \alpha,
\end{equation}
where $\alpha =4  \Gamma_l \Gamma_r / [\overline{\Gamma}( \Gamma_l + \Gamma_r )]$ fluctuates from peak to peak with statistics that depend on $\beta$, $\Sigma$, and $s$, and $\chi_{s=1}=1/8$ and $\chi_{s=2} = 3 - 2\sqrt{2}$ account for effects of Kramers degeneracy on Coulomb blockade \cite{Aleiner:2002in}.
At zero magnetic field, the distribution of $\alpha$ for weak spin-orbit coupling is given by \cite{Jalabert:1992hb, Aleiner:2002in, Ahmadian:2006jx}
\begin{equation}\label{eq:pbeta1}
P_{\beta=1,\Sigma=1,s=2}( \alpha ) = \sqrt{ \frac{1}{ \pi \alpha }} e^{-\alpha}, 
\end{equation}
whereas for strong spin-orbit coupling it is given by \begin{equation} \label{eq:pbeta4}
P_{\beta=4,\Sigma=1,s=2}( \alpha ) = 16 \alpha^3 e^{-2 \alpha} \left( K_0( 2 \alpha ) + \left( 1 + \frac{1}{4 \alpha} \right) K_1( 2 \alpha ) \right),
\end{equation}
where $K_0$ and $K_1$ are modified Bessel functions.
The distributions have $\overline{\alpha}=1/2$ and $\overline{\alpha}=4/5$ for weak and strong spin-orbit coupling, respectively.

A consequence of the equality 
\begin{equation}\label{eq:pbeta2} 
P_{\beta=2, \Sigma=2, s=1}( \alpha ) = P_{\beta=4, \Sigma=1, s=2} (\alpha) 
\end{equation}
is that for strong spin-orbit coupling, the peak height distribution at high field is expected to be a scaled version of the zero-field distribution, {\em decreased} by the ratio $\chi_{s=2}/\chi_{s=1}\sim 1.4$ \cite{Ahmadian:2006jx}. This is in contrast to the weak spin-orbit case, where the high-field distribution differs markedly in shape from the zero-field distribution, and the high-field mean height is {\em increased} by a factor of 4/3 compared to zero field \cite{Jalabert:1992hb}, consistent with experiment \cite{Chang:1996aa, Folk:1996gy}.

\begin{figure}[t]
\includegraphics{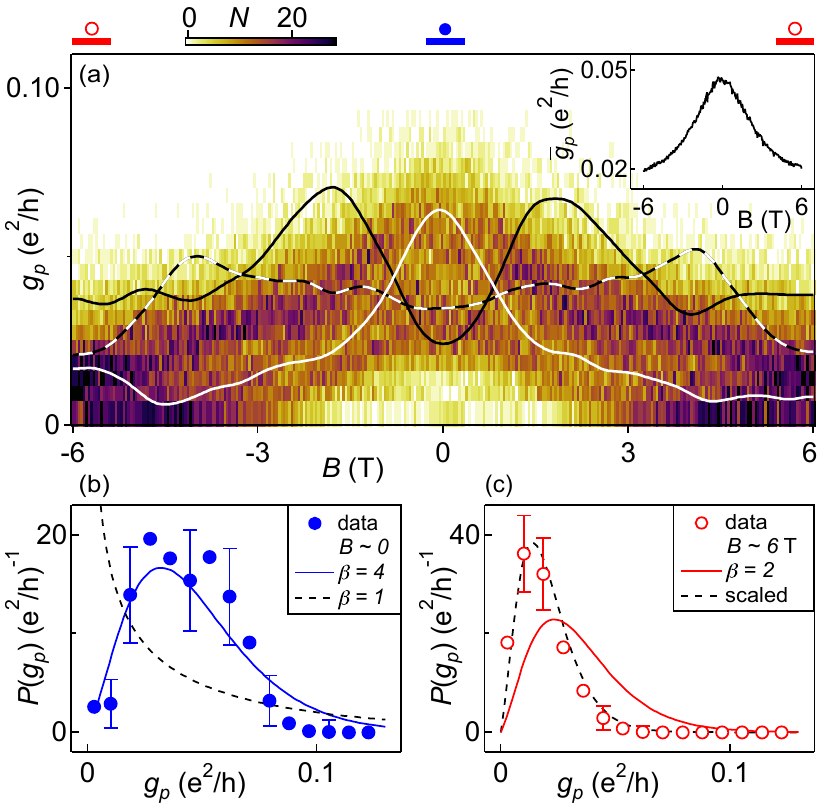}
\caption{\label{fig:3}
(a) Histograms of Coulomb blockade peak heights (color scale) as a function of magnetic field, $B$. Line traces show the smoothed conductance of three individual Coulomb peaks. Average peak height decreases with $B$, while individual peak heights fluctuate.
\textit{Inset:} Measured mean peak height, $\overline{g}_p$, as a function of $B$, extracted from data in main figure.
(b) Peak height distribution, $P(g_p)$, for $B \sim 0$ (range shown as blue band at the top of (a)).
Theory curves from Eq.~(\ref{eq:pbeta4}) (solid) and Eq.~(\ref{eq:pbeta1}) (dashed).
(c) Peak height distribution, $P(g_p)$, for $|B| \sim 6$~T (range shown as red bands at the top of (a)).
Theory curves from Eq.~(\ref{eq:pbeta2}) (solid), which is the same as Eq.~(\ref{eq:pbeta4}) scaled by $8(3-2\sqrt{2})\sim 1.4$ and Eq.~(\ref{eq:pbeta4}) scaled by a factor of 2.3 (dashed). 
The single experimental parameter $\overline{\Gamma}/(kT)$ is fixed using $\overline{g}_p$ at $B=0$ from (a) inset.
}
\end{figure}

 The measured device was formed from a Ge-Si core-shell nanowire (10~nm Ge core, 2~nm Si shell)
\footnote{The device shown in Fig.~\ref{fig:1} is lithographically identical to the one studied here. However, in the measured device the nanowire forms a smaller angle with $B$ of $31^{\circ}$ determined from optical microscopy.}
placed on an array of Cr/Au  bottom gates (2 nm/20 nm thick, 20 nm wide, 60 nm pitch) patterned by electron beam lithography on a native-oxide Si wafer, then covered with 25 nm of HfO$_2$ (grown by atomic layer deposition at $180~^{\circ}\mathrm{C}$) before depositing the wires. Patterned
Ti/Pd ohmic contacts were deposited following a 3 s buffered HF etch. Conductance was measured in a dilution refrigerator with electron temperature $T \sim 100~\mathrm{mK}$ using standard lock-in techniques with ac excitation $V_{ac}=100\, \mathrm{\mu V}$, except where noted.
The lock-in excitation was chosen to be as large as possible without altering the peak height distribution.
An in-line resistance of $4.2~\mathrm{k \Omega}$ was subtracted from all data.

\begin{figure}
\includegraphics{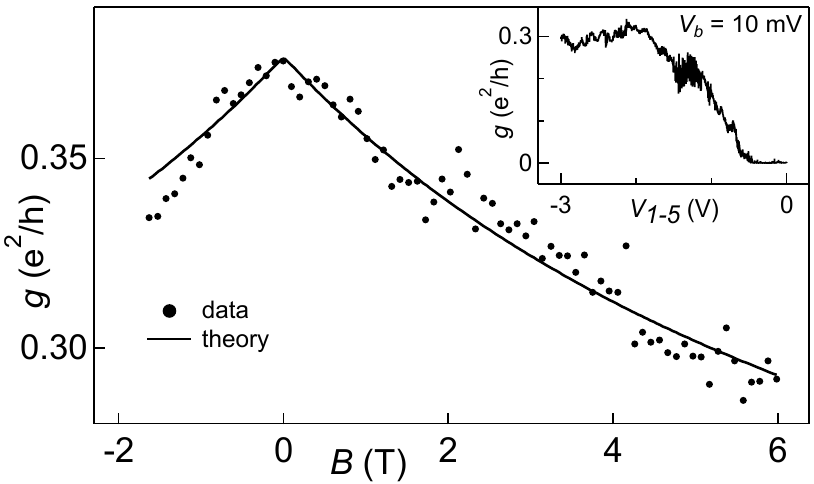}
\caption{\label{fig:4}
Two-terminal conductance, $g$, in the open-wire regime as a function of magnetic field, $B$ (points) along with  theory, based on Eq.~(\ref{eq:biganti}), including contact resistance.
Fit bounds spin-orbit length, $l_\mathrm{so} < 20~\mathrm{nm}$.
\textit{Inset:} Pinch-off curve at bias $V_b = 10~\mathrm{mV}$. Saturation at $g\sim0.3~\mathrm{e^2/h}$ indicates decreasing mobility in the open regime (see text).
}
\end{figure}

A typical orbital level spacing of $\Delta \sim 0.2~\mathrm{meV}$ was measured from Coulomb blockade spectroscopy, as shown in Fig.~1, inset.  The number of holes, $N_H$, in the Coulomb blockade regime was estimated to be roughly 600, based on counting Coulomb oscillations. The length of the quantum dot was in the range $L=200-600~\mathrm{nm}$, corresponding to the length of the middle segment and the wire.
For wire width $w = 10$ nm, this gives $M=4 w/\lambda_F=4-6$ occupied transverse modes, using a 3D estimate for the Fermi wavelength, $\lambda_F= ( 2 \pi^2 L w^2 / 3 N_H )^{1/3} \sim 6-9~\mathrm{nm}$.
Elastic scattering length $l=h \mu / \lambda_F e = 35 - 50~\mathrm{nm}$ and mobility $\mu \sim 800~\mathrm{cm^2/V s}$ were extracted from the slope of the pinch-off curve (Fig.~\ref{fig:4}, inset) using $g = (\pi w^2/4 L )\mu n e$ \cite{Lu:2005jx,Hao:2010di}. Values in the open regime differ somewhat, as discussed below.

Figure~\ref{fig:1} shows two-terminal conductance of the nanowire as a function of a common voltage on gates 2, 3 and 4, denoted $V_{2-4}$, for a common voltage on gates 1 and 5, $V_{1,5}$, corresponding to open (top trace) and tunneling (bottom trace) regimes.
The open regime showed weak dependence on gate voltage, with an onset of Coulomb oscillations as conductance decreases; the tunneling regime showed well defined Coulomb blockade peaks with fluctuating heights. The heights of neighboring peaks appear correlated over roughly two peaks, even at the lowest temperatures, similar to \cite{Folk:1996gy}, which decreases the effective ensemble size.

Representative sets of Coulomb blockade peaks at $B \sim 0$ and $6~\mathrm{T}$ [Figs.~2(a,b)] show a decrease in average peak height at high field, as expected for strong spin-orbit coupling.
As temperature was increased above $\Delta$, fluctuations in peak height decreased rapidly, consistent with a simple model that assumes resonant transport through multiple, uniformly spaced levels \cite{*[{``Picket fence" model in: }] [{.}]  Alhassid:1998bv} [Figure~\ref{fig:2}(c)]. 
Note that the same ensemble was used for each temperature. This presumably accounts for the correlated departures from theory at low temperature.  The discrepancy with theory at high temperature is unexplained, and is reminiscent of \cite{Patel:1998jy}.

Peak height histograms for $m = 142$ Coulomb peaks (see Supplemental Material for gate voltage ranges) show the evolution of the distribution as a function of magnetic field [Fig.~3(a)]. 
The observed decreasing average peak height at higher fields---Coulomb blockade antilocalization---as well as the maximum in the distribution away from zero height at all fields, are both signatures of strong spin-orbit coupling.

Figures 3(b,c) show peak height distributions, ${P( g_p ) =  ( \overline{\alpha}/\overline{g}_p )P_{\beta,\Sigma,s} ( \overline{\alpha} g_p /\overline{g}_p ) = N / ( m W)} $, where $W$ is the bin width and $N$ is the bin count in Fig.~3(a), at low and high magnetic fields.

The low-field data in Fig.~3(b) agree with the theoretical distribution for strong spin-orbit coupling ($\beta = 4$), with the mean peak height taken from Fig.~3(a), and are inconsistent with the theoretical distribution for weak spin-orbit coupling ($\beta = 1$).
The high-field data in Fig.~3(c) are consistent with a scaled version of the low-field theoretical distribution, as expected for strong spin-orbit coupling, but with a scale factor of $\sim 2.3$ rather than the theoretically predicted factor of 1.4.
The reason for this discrepancy---qualitative scaling, but not by the predicted factor---is not understood, but may result from changes in tunnel rates out of the dot or changes in density of states in the leads, which are also segments of nanowire.

To compare antilocalization in the Coulomb blockade regime to the open-wire regime, we tuned the device to more negative gate voltages, where Coulomb blockade oscillations were absent [see Fig.~\ref{fig:1}(a)]. The number of holes was larger in the open regime,  $N_H \sim 1700$ and $\lambda_F \sim 6~\mathrm{nm}$, again determined by counting Coulomb oscillations and assuming the device is depleted at pinch-off.
The inset of Fig.~\ref{fig:4} shows that high-bias conductance saturates at larger negative gate voltages, indicating a decreasing mobility with increasing density. Similar behavior has been reported in Ge-Si nanowires \cite{Lu:2005jx}, Ge nanowires \cite{Wang:2003fj}, and Si heterostructures \cite{Cheng:1973vc}.
In Si heterostructures, this decrease in mobility was explained as resulting from carriers being pulled toward the rough heterointerface, as well as an increase in phase space for scattering as more transverse subbands are occupied \cite{Cheng:1973vc,Mori:1979dp}. Presumably, comparable effects occur in wires. 

Magnetoconductance, $g(B)$, measured in the open-wire regime, is shown in Fig.~\ref{fig:4} along with a theory curve that includes contributions from the wire, $g_{w}(B)^{-1}$, as well as from the two contacts, each set to $g_{c}^{-1} = 2e^{2}/h$ near the onset of Coulomb blockade, $g(B) = [2 g_c^{-1} + g_w(B)^{-1} ]^{-1}$. Following Ref. \cite{Kurdak:1992cp}, we use the expression 
\begin{align}
g_w(B) = g_{\infty} - \frac{2 e^2}{h} \frac{1}{L} \biggl[ \label{eq:biganti}
&\frac{3}{2} \biggl( \frac{1}{D \tau_\phi} + \frac{4}{3 D \tau_\mathrm{so}} + \frac{1}{D \tau_B} \biggr)^{-1/2} \\ \nonumber
-&\frac{1}{2} \biggl( \frac{1}{D \tau_\phi} + \frac{1}{D \tau_B} \biggr)^{-1/2} \\ \nonumber
-&\frac{3}{2} \biggl( \frac{1}{D \tau_\phi} + \frac{4}{3 D \tau_\mathrm{so}} + \frac{1}{D \tau_e} +  \frac{1}{D \tau_B} \biggr)^{-1/2} \\ \nonumber
+&\frac{1}{2} \biggl( \frac{1}{D \tau_\phi} + \frac{1}{ D \tau_e} + \frac{1}{D \tau_B} \biggr)^{-1/2}
\biggr],
\end{align}
for the magnetoconductance of the wire, where $g_{\infty}$ is the classical (background) conductance, $L\sim600~\mathrm{nm}$ is the length of the occupied region of the nanowire, $D$ is the diffusion constant, and $\tau_\phi$, $\tau_\mathrm{so}$, $\tau_B$, $\tau_e$ are the dephasing, spin relaxation, magnetic, and impurity-impurity scattering times.
We note that in the present study, where $l_e \ll l_\phi$, the last two terms of Eq.~(5) do not play an important role, and in principle could be dropped.
We retain these terms, though they have no discernible effect on the fits, for consistency with the existing literature \cite{Hansen:2005ek,Hao:2010di,Kallaher:2010iq,Kallaher:2010uh} for $w < l_e$.

The transport scattering length, $l_{t}=2D/v_{f}$, where $v_f$ is the Fermi velocity, the dephasing length, $l_\phi$, and the spin precession length, $l_\mathrm{so}$, then appear as \cite{Beenakker:1988dv,Kettemann:2007eg}
$D \tau_\phi =  l_\phi^2/2$, 
$D \tau_e = l_t l_{e}/2$, 
$D \tau_B = C_1\,l_t l_B^4/w^3 + C_2\,l_t l_{e} l_B^2/w^2$, 
and $D \tau_\mathrm{so} = C_3\, l_t l_\mathrm{so}^4/w^3$, 
where $l_B^{2}= \hbar/e B$. Contants $C_1=4 \pi \, (9.5)$, $C_2 = 3 \, (24/5)$ apply for diffusive (specular) boundary scattering  \cite{Beenakker:1988dv}, and we interpolate between these values for specularity, $\epsilon$, between zero (fully diffusive) and one (fully specular). 
We use the specular value $C_3 = 130$ \cite{Kettemann:2007eg}, lacking a theoretical value for diffusive boundary scattering.  
The ratio of scattering lengths depends on specularity and sample width, $l_t/l_{e} = F( w/l_e, \epsilon )$, with $F(\cdot \, , 1)$ = 1 
\footnote{ $F$ is given by \cite{Dingle:1950ew}
\[
\begin{split}
F( \kappa, \epsilon ) = &1 - \frac{12}{\pi} (1 - \epsilon^2) \\
&\sum_{\nu=1}^\infty \nu \epsilon^{\nu-1} \int_{0}^1 dx \sqrt{ 1 - x^2 } S_4( \nu \kappa x ), 
\end{split}
\]
where $S_4( \lambda ) = \int_0^{\pi/2} d\theta e^{-\lambda/\sin \theta} \cos^2 \theta \sin \theta$.}.
These expressions require for $\lambda_F < w$ and $w < l_e$, the former barely satisfied for $\lambda_F = 6~\mathrm{nm}$.

Four free parameters, $l_\mathrm{so}$, $g_\infty$, $l_e$, and $l_\phi$, are used to fit theory to data. The transport scattering length is found from $l_t =  (4 L / \pi w^2 ) h g_\infty / \lambda_F n e^2$, where $n= 4 N_H / \pi w^2 L$ is the 3D hole density (a reasonable model, given six occupied transverse modes). Specularity can then be found by inverting $l_t /l_{e} = F( w/l_e, \epsilon )$, and the Fermi wavelength can be found from the 3D density, $\lambda_F = ( 8 \pi / 3 n )^{1/3}$. As seen in Fig.~4, the model fits the data very well, and gives the following ranges for transport parameters, $g_\infty = 0.2 - 0.7~\mathrm{e^2/h}$, $l_e < 10~\mathrm{\mu m}$, $l_t = 15$-$25~\mathrm{nm}$
$l_\phi = (0.2-1.2)~\mathrm{\mu m}$, specularity in the range $\epsilon = 0.4-1$, and $l_\mathrm{so} < 20~\mathrm{nm}$. (Allowing $l_\mathrm{so} > 20~\mathrm{nm}$ gives good fits only with $l_e > 10~\mathrm{ \mu m}$, which we rule out as unphysical.) 

As a comparison between open and nearly isolated regimes, we note that the observation of antilocalization in Coulomb blockade implies $\epsilon_\mathrm{so} > \Delta$ where $\epsilon_{\mathrm{so}}$ is the spin-orbit energy in the dot \cite{Ahmadian:2006jx}.
To convert this into a spin-orbit length we assume the simple relation $\epsilon_{\mathrm{so}} = \hbar^2 / ( 2 m^{*} l_{\mathrm{so}}^2 )$ \cite{Kloeffel:2011bg} and the bulk heavy-hole effective mass $m^{*} = 0.28~m_e$. This gives $l_\mathrm{so} < 25~\mathrm{nm}$, consistent with the open regime measurement of $l_\mathrm{so} < 20~\mathrm{nm}$.

It is interesting to consider the reason for the large magnetic field scale associated with antilocalization in both regimes. Flux cancellation due to boundary scattering is known to enhance the effective magnetic length \cite{Beenakker:1988dv}. Flux cancellations of the effective spin-orbit magnetic field also occur \cite{*[{ Cancellation occurs only for Rashba spin-orbit, expected in Ge/Si: }] [{.}]  Kettemann:2007eg}. These effects roughly cancel out, and the field scale for antilocalization is then $l_B = l_\mathrm{so}$, or $B^* = \hbar / ( e l_\mathrm{so}^2 )=3~\mathrm{T}$ for $l_\mathrm{so}=15~\mathrm{nm}$. 

In summary, we have presented an experimental study of Coulomb blockade peak height statistics in a Ge-Si nanowire. Peak height distributions as well as the field dependence of average peak height (antilocalization) are consistent with the effects of strong spin-orbit coupling. However, the observed decrease in average peak height with applied magnetic field is larger than expected. Magnetoconductance of the same device configured as an open wire yields consistent results.
Further investigation of the spin-orbit strength in this system could come from spectroscopic measurements of orbital anticrossings in a quantum-dot, or from electric-dipole spin resonance measurements in a Ge-Si double quantum dot. Combined with the expectation of long spin dephasing times in Ge-Si quantum dots, the strong spin-orbit coupling found in this work makes Ge-Si nanowire quantum dots attractive for spin qubit applications.

We thank Igor Aleiner for valuable discussions. Research supported by the Danish National Research Foundation,  The Office of Science at the Department of Energy, the National Science Foundation (PHY-1104528), and the Defense Advanced Research Projects Agency through the QuEST Program.

\bibliography{main}

\end{document}